\documentclass[prl,twocolumn,showpacs,preprintnumbers,amsmath,amssymb,superscriptaddressm]{revtex4}

\usepackage{graphicx}% Include figure files
\usepackage{dcolumn}% Align table columns on decimal point
\usepackage{bm}% bold math

\begin{document}

\title{Crossover from Coulomb blockade to quantum Hall effect in suspended graphene nanoribbons}

\author{Dong-Keun Ki}
\author{Alberto F. Morpurgo}
\email{Alberto.Morpurgo@unige.ch} \affiliation{D\'{e}partment de
Physique de la Mati\'{e}re Condens\'{e}e (DPMC) and Group of Applied
Physics (GAP), University of Geneva, 24 Quai Ernest-Ansermet, CH1211
Gen\'{e}ve 4, Switzerland}

\date{\today}

\begin{abstract}
Suspended graphene nano-ribbons formed during current annealing of
suspended graphene flakes have been investigated experimentally.
Transport measurements show the opening of a transport gap around
charge neutrality due to the formation of "Coulomb islands",
coexisting with quantum Hall conductance plateaus appearing at
moderate values of magnetic field $B$. Upon increasing $B$, the
transport gap is rapidly suppressed, and is taken over by a much
larger energy gap due to electronic correlations. Our observations
show that suspended nano-ribbons allow the investigation of
phenomena that could not so far be accessed in ribbons on SiO$_2$
substrates.

\end{abstract}
\pacs{72.80.Vp,73.43.Qt,73.23.Hk,85.35.-p}

\maketitle

Theoretical studies indicate that in graphene nano-ribbons (GNRs)
--long and narrow graphene channels-- Dirac electrons should give
origin to a number of unique phenomena, accessible in transport
experiments. Examples include the controlled opening of a band
gap~\cite{Nakada96}, the realization of gate-tunable magnetic
states~\cite{Son06}, or the interplay between Landau and size
quantization ~\cite{Peres06}. Theories, however, usually consider
disorder-free GNRs with idealized edge
structure~\cite{Nakada96,Son06,Peres06}, whereas disorder is
invariably present in experimental systems, which leads to the
formation of localized states in a wide energy
range~\cite{Han07,Moliter09,Oostinga10}. In the presence of strong
disorder, transport through GNRs near charge neutrality is mediated
by hopping, and GNRs behave as an ensemble of randomly assembled
quantum dots, resulting in the opening of a so-called transport
gap~\cite{Han07,Moliter09,Oostinga10}. In this regime, even the most
basic manifestations of the Dirac character of electrons, such as
the half-integer conductance quantization in a magnetic
field~\cite{Novoselov05}, are washed
out~\cite{Moliter09,Oostinga10}.

A complete elimination of disorder would  require control of the
edge structure ~\cite{Poumirol10,Jiao11}, and the fabrication of
suspended GNR devices to avoid the influence of substrate-induced
disorder~\cite{Du08}. The high electronic quality of suspended
graphene devices has indeed been demonstrated --among others-- by
the observation of ballistic transport~\cite{Du08,Tombros11} and of
the fractional quantum Hall effect in
monolayers~\cite{Du09,Bolotin09}, and of small interaction induced
gaps in bilayers~\cite{Weitz10}. Crucial to achieve high-quality is
a current annealing step, i.e., forcing a large current to heat
graphene, in order to desorb adsorbates from the
surface~\cite{Du08,Moser07}. The large current required can lead to
cleavage of graphene, which usually results in device failure. We
have found that in several cases a partial graphene cleavage can
occur~\cite{Tombros11,Moser09}, leading to the formation of narrow
suspended nano-ribbons.

Here, we report on magneto-transport measurements through a device
produced in this way. We find that, while a disorder-induced
transport gap still opens around charge neutrality (as observed in
lithographically defined GNRs on SiO$_2$
substrates~\cite{Han07,Moliter09,Oostinga10}), Coulomb blockade is
strongly suppressed by the application of only a moderate magnetic
field ($B \approx 2$ T)~\cite{Oostinga10,Poumirol10}. This leads to
quantum Hall conductance plateaus at finite carrier density, and the
occurrence of a strongly insulating state (with a gap increasing
with magnetic field) around charge neutrality~\cite{Checkelsky08}.
These phenomena, normally observed only in sufficiently clean
graphene ~\cite{Du08,Tombros11,Du09,Bolotin09}, indicate that
suspension of GNRs result in a significant improvement of their
electronic quality, enabling the observation of phenomena not
accessible in GNRs on SiO$_2$
substrates~\cite{Han07,Moliter09,Oostinga10,Poumirol10}.

\begin{figure}[t]
\includegraphics[width=8.5cm]{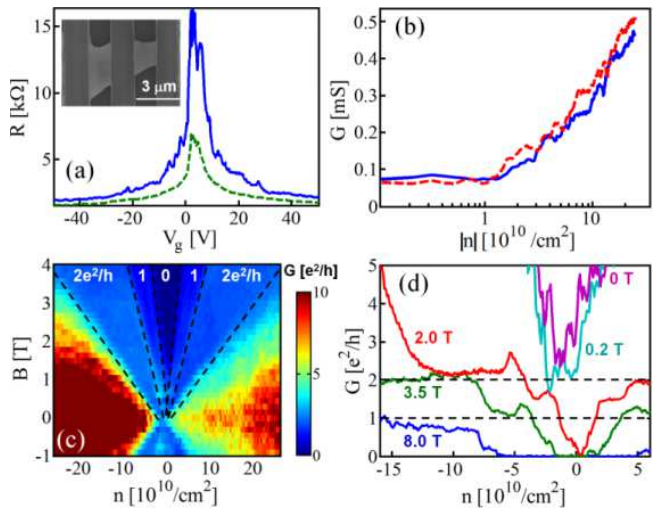}
\caption{\label{fig1} (Color online) (a) Two-terminal resistance
($R$) versus $V_g$ of the wide suspended graphene device shown in
the inset, measured at $T=0.24$ K (blue line) and 2.0 K (Green
dashed line). (b) Semi-log plot of the conductance $G$ measured at
0.24 K, as a function of carrier density $n$ for both electrons
(blue solid line) and holes (red dashed line). The resistance peak
width is approximately $10^{10}$ carriers/cm$^2$.(c) Conductance $G$
as a function of carrier density $n$ and magnetic field $B$, showing
well defined quantum Hall states with $G=1$ and 2$e^2/h$, and an
insulating state around charge neutrality. (d) $G$($n$) at different
values of $B$ selected from (c), zooming in on quantum Hall
plateaus; the dashed lines correspond to $G=1$ and 2$e^2/h$.}
\end{figure}

Wide suspended graphene devices are fabricated using
polydimethylglutarimide (PMGI) resist (LOR, MicroChem) as a
sacrificial supporting layer, as described by Tombros {\it et al.}
(see Ref. 18 for details). Measurements are performed in a
two-terminal configuration with Ti/Au contacts (10/70 nm) separated
by a distance of approximately 0.5-1 $\mu$m (depending on the
device), using the highly doped Si substrate as gate electrode
(approximately 1 $\mu$m away from graphene). The inset in Fig.
\ref{fig1}(a) shows a scanning electron microscope (SEM) image of a
successfully annealed device.

\begin{figure}[t]
\includegraphics[width=8.5cm]{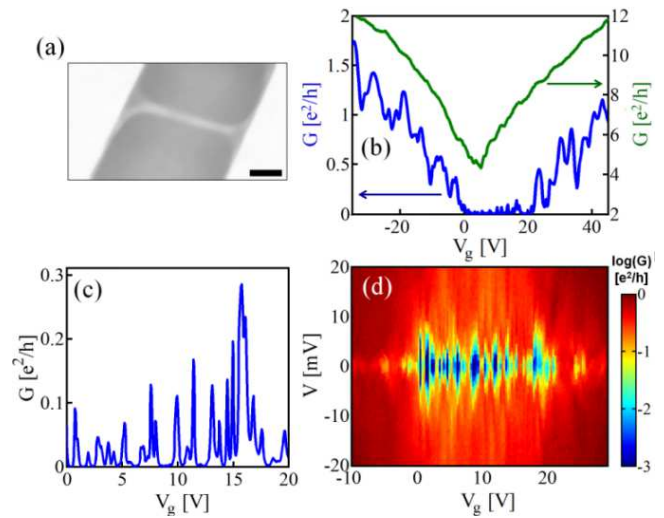}
\caption{\label{fig2} (Color online) (a) SEM image of a GNR formed
while annealing a suspended graphene layer (the bar is 100 nm long).
(b) Gate dependence of the differential conductance
$G=\frac{dI}{dV}(V_g)$ measured after a first annealing step of a
suspended graphene device (green line on top) and after a second
step (blue line at the bottom), in which a suspended GNR is formed.
Panel (c) zooms in on the transport gap region of (b), and shows
resonances due to Coulomb blockade. (d) log($G$) as a function of
bias $V$ and gate $V_g$ voltages, showing the Coulomb diamonds
characteristically seen in GNRs. All data are taken at 4.2 K. }
\end{figure}

\begin{figure}[t]
\includegraphics[width=8.5cm]{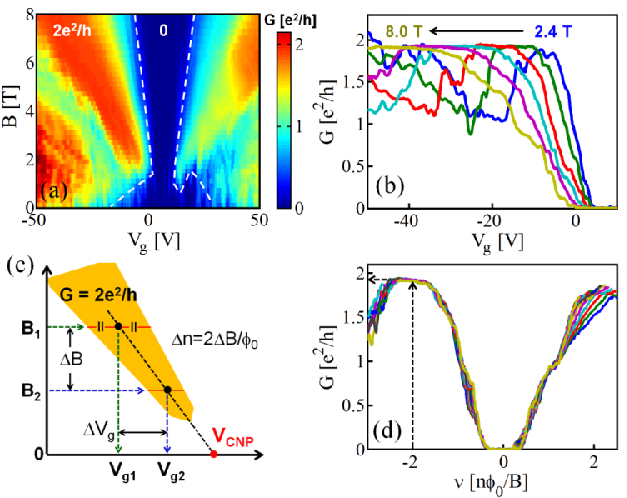}
\caption{\label{fig3} (Color online) (a) Color plot of
$G$($V_g$,$B$), with quantized conductance values indicated. The
broken line is guide to the eye, separating the conducting and
insulating regions as a function of $V_g$ and $B$. (b) $G$($V_g$) at
$B=2.4$ T, 3.4 T, 4.4 T, 5.4 T, 6.5 T, and 8.0 T (from the right to
the left, respectively), showing conductance plateaus for hole
accumulation. (c) Schematic view of the procedure followed to
estimate $V_{CNP}$ and $\alpha$($\equiv\Delta n/\Delta V_g$)
described in the text. (d) $G$ as a function of the filling factor
$\nu$($\equiv n\phi_0/B$) in the magnetic field range between 5.0
and 8.0 T: all different measurements tend to collapse on a single
curve. The presence of 2$e^2/h$ conductance plateaus for electron
accumulation becomes apparent in this plot. All data are taken at
4.2 K.}
\end{figure}

Current annealing was performed by slowly ramping up the bias
voltage $V$  to $\sim1$-2 V (depending on the specific sample) in
vacuum at 4.2 K, and maintaining the applied bias for extended
periods of time. For most devices, this procedure had to be repeated
several times before a sharp resistance peak at approximately zero
gate voltage ($V_g=0$ V) could be observed (see Figs.
\ref{fig1}(a),(b)). In successfully annealed devices, low-field
integer quantum Hall effect with a complete lifting of spin and
valley degeneracy was observed, together with a  strongly insulating
state appearing at charge neutrality upon the application of a
magnetic field $B$ (see Figs. \ref{fig1}(c),(d)). These observations
and the narrow width of the resistance peak around charge neutrality
($\simeq 10^{10}$ cm$^{-2}$) indicate the high device
quality~\cite{Du08,Tombros11,Du09,Bolotin09}.

In several different devices, we observed sudden, large increases in
resistance during annealing, originating from partial breaking of
the graphene layer (see, e.g., Fig. \ref{fig2}(a)). In these cases,
the measured $V_g$ dependence of the conductance $G$ exhibits
features that are characteristically observed in
nano-ribbons~\cite{Han07,Moliter09,Oostinga10}. Figs.
\ref{fig2}(b)-(d) show data from a device in which the phenomenon
was seen more clearly. Specifically Fig. \ref{fig2}(b) shows the
$V_g$ dependence of the conductance before (green curve) and after
(blue curve) a final annealing step, which resulted in the opening
of a transport gap in the $V_g$ range between 0 and 20 V. In this
range, a random sequence of conductance peaks is observed (Fig.
\ref{fig2}(c)), and measurements of the differential conductance
versus $V$ and $V_g$ show "Coulomb diamonds" characteristic of a
disordered array of quantum dots (Fig. \ref{fig2}(d)). This is the
same behavior observed in GNRs on SiO$_2$
substrates~\cite{Han07,Moliter09,Oostinga10}.

Measurements as a function of $B$ illustrate the difference between
suspended GNRs and GNRs on SiO$_2$
~\cite{Moliter09,Oostinga10,Poumirol10}. Fig. \ref{fig3}(a) shows
the $V_g$ and $B$ dependence of the conductance, where quantum Hall
states with $G=2 e^2/h$ are observed for holes and for electrons.
The quantization of the conductance is better seen in Fig.
\ref{fig3}(b), which shows cuts at fixed $B$ of the color plot in
Fig. \ref{fig3}(a). For negative $V_g$, the plateaus appear at
$B\simeq 2-2.5$ T, whose conductance is very close to 2$e^2/h$
($\sim 1.9 e^2/h$; the deviation is due to a 350 $\Omega$ contact
resistance) as expected for Dirac electrons~\cite{Novoselov05}. For
positive gate voltages, the plateaus become well defined only at
larger magnetic field (see Fig. \ref{fig3}(d))~\cite{note1}. Around
charge neutrality and for $B>2$ T, a strongly insulating state is
seen, similarly to what is normally observed in large,  clean
suspended graphene flakes~\cite{Du09,Bolotin09}, originating from
interaction-induced electronic correlations~\cite{Checkelsky08}.

To further test the quantum Hall nature of the plateaus we plot the
conductance $G$ as a function of filling factor $\nu = n \phi_0/B$
(with $\phi_0 = h/e$ and $n$ carrier density), to check if all data
collapse on a single curve as expected. To extract the carrier
density $n$ as a function of  $V_g$, we exploit the fact that at the
center of a 2$e^2/h$ conductance plateau at a fixed magnetic field
$B$, $\nu=2$ (owing to spin degeneracy) so that $n=2\phi_0/B$. We
then find that the carrier density scales linearly with gate
voltage, as it should. The slope $\alpha = \Delta n/\Delta V_g=7.8
\times 10^9$ cm$^{-2}$V$^{-1}$ is slightly larger than the value,
4.7 $\times 10^9$ cm$^{-2}$V$^{-1}$, estimated from the formula for
a parallel plate capacitor, as expected, since the ribbon
width~\cite{note1} is smaller than the distance to the gate (the
value of $\alpha$ agrees quantitatively with a recent study of
suspended graphene in the quantum Hall regime~\cite{Marun11}, where
an enhanced capacitance at the edges was found). By extrapolating to
$B=0$ T, we find that charge neutrality occurs at $V_g=V_{CNP}=8.5$
V (see Fig. \ref{fig3}(c)). Using this relation between $n$ and
$V_g$, we plot the conductance as a function of $\nu$ in the $B$
range between 5.0 and 8.0 T (Fig. \ref{fig3}(d)). On the hole side
all curves collapse together nearly perfectly; on the electron side
the trend is similar, and although the collapse is not as good, the
presence of a 2$e^2/h$ quantum Hall plateau   becomes apparent.

\begin{figure}[t]
\includegraphics[width=8.5cm]{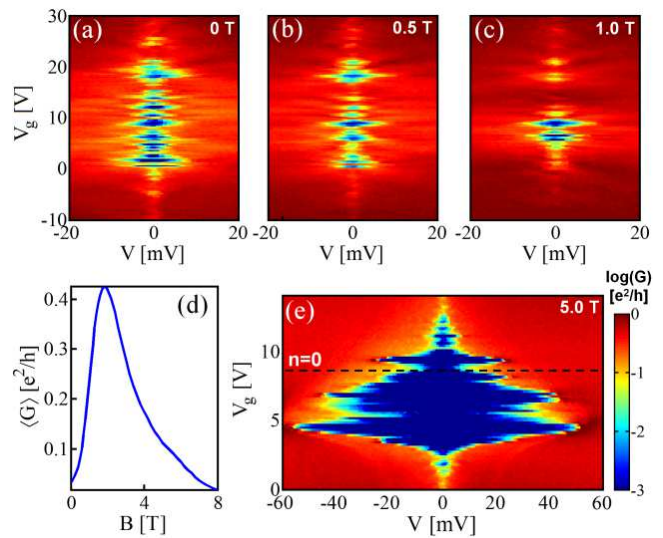}
\caption{\label{fig4} (Color online) (a-c) From left to right,
logarithm of the differential conductance $G=\frac{dI}{dV}$ versus
$V$ and $V_g$, for $B=0$ T, 0.5 T, and 1 T: increasing $B$ causes a
decrease of the $V_g$ interval in which Coulomb diamonds dominate
low-bias transport. (d) Conductance averaged in $V_g$'s between 0 V
and 20 V, $\langle G \rangle$, as a function of $B$. At $B>2$ T,
$\langle G \rangle$ decreases due to the large, correlation-induced
gap around the charge neutrality, which becomes larger as $B$
increases. (e) log($G$) versus $V$ and $V_g$ at $B$=5.0 T, showing
the opening of a large gap around charge neutrality due to
electronic correlations (the color scale in the legend is the same
for all panels). The broken line at $V_g=8.5$ V corresponds to
$V_{CNP}$ extracted from the quantum Hall effect analysis (see Fig.
\ref{fig3}(c)). All data are taken at 4.2 K}
\end{figure}

We conclude that in suspended GNRs the transport gap due to Coulomb
islands at $B=0$ T coexists, at $B>2$ T, with quantum Hall effect at
finite carrier density, and with a large-gap correlated state around
charge neutrality. As these phenomena occur at low magnetic field
only when the level of disorder is sufficiently
low~\cite{Du08,Tombros11,Du09,Bolotin09}, their observation
indicates that the quality of suspended GNRs created during current
annealing is considerably better than that of GNRs on SiO$_2$
substrates~\cite{Han07,Moliter09,Oostinga10,Ribeiro11}. Similarly to
larger suspended graphene devices, the low disorder level in
suspended GNRs is due to the absence of a substrate and to the
current annealing. The current-induced cleaving process --which
occurs in vacuum-- may also lead to better edges~\cite{Jia09} as
compared to lithographically defined
GNRs~\cite{Han07,Moliter09,Oostinga10}.

The higher quality enables the investigation of  the competition
between Coulomb blockade --originating from geometrical confinement
in the presence of disorder-- and the quantum-Hall effect, due to
magnetic confinement. One interesting aspect is the dependence of
the transport gap on magnetic field. Fig. \ref{fig3}(a) shows that
the transport gap due to Coulomb islands (i.e., the $V_g$ interval
where the conductance is strongly suppressed) narrows down with
increasing $B$ from 0 to 1 T. The phenomenon is illustrated in more
detail in Figs. \ref{fig4}(a)-(c), which show measurements of
Coulomb diamonds for different values of $B$. While at $B=0$ T
diamonds appear when  $V_g$ is between 0 and $\simeq 20$ V, at $B=1$
T they are only present  between $\simeq 5$ and 10 V.
Correspondingly, the conductance $\langle G \rangle$ averaged over
$V_g$  between 0 V and 20 V increases by more than ten times as $B$
is increased from 0 to $\sim2$ T (Fig. \ref{fig4}(d)). An increase
in the averaged conductance was found previously in GNRs on SiO$_2$
substrates~\cite{Oostinga10,Poumirol10}, and correctly attributed to
a  magnetic-field induced increase of the electron localization
length, but a full crossover from a Coulomb blockaded regime to
fully developed conductance quantization has not been observed
previously~\cite{Moliter09,Oostinga10,Poumirol10,Ribeiro11}.

Note that, at low carrier density, a large increase in the
two-terminal conductance of a narrow disordered wire (i.e. where at
$B=0$ T electron states are localized) upon entering the quantum
Hall regime is a manifestation of the Dirac fermion character of the
charge carriers. If electrons were described by the Schrodinger
equation --as it would be the case for a narrow wire defined in a
conventional two-dimensional electron gas-- entering the quantum
Hall regime at low carrier density would result in a so-called
sub-band depopulation (due to the enhanced magnetic field
confinement), and in a decrease of the two terminal conductance with
increasing $B$~\cite{Beenakker91}. For Dirac electrons, the
existence of a zero-energy Landau level is responsible for this
difference, and causes the conductance to increase.

\begin{figure}[t]
\includegraphics[width=8.5cm]{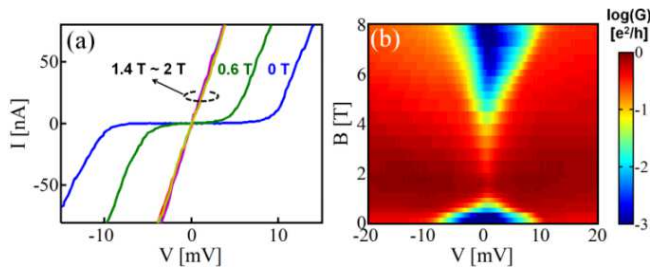}
\caption{\label{fig5} (Color online) (a) $I-V$ characteristics of
the device at $V_g=1.4$ V (at $B=0$ T, 0.6 T and in the range
between 1.4 T and 2.0 T, as indicated). (b) Logarithm of the
differential conductance $G=\frac{dI}{dV}$ versus $V$ and $B$ for
$V_g=1.4$ V away from the charge neutrality, showing the closing of
the transport-gap with increasing $B$ (the lo-bias conductance
suppression above 4 T is due to the insulating state around the
charge neutrality). All data are taken at 4.2 K.}
\end{figure}

The crossover from the Coulomb-blockaded regime at $B=0$ T
--characterized by non-Ohmic $I-V$ curves-- to a linear transport
regime at finite magnetic field $B$ (see Fig. \ref{fig5}(a)) is also
apparent in measurements at fixed $V_g$ in the "transport gap"
range. Fig. \ref{fig5}(b) shows data taken at $V_g=1.4$ V, in which
a 10 meV gap due to charging energy at $B=0$ T closes when $B$ is
increased to 1.4 T (at $B>4$ T, $G$ decreases again due to the large
gap associated to the correlated state around the charge neutrality
point). The full suppression of Coulomb blockade is expected in GNRs
exhibiting a clear quantum-Hall effect, since plateaus with
quantized conductance (Fig. \ref{fig3}) indicate that transport is
mediated by edge states and that backscattering is absent. This
implies that electrons are delocalized over the full device length,
so that transport is not any more  blocked by the Coulomb charging
energy associated to localized states in "disorder-induced islands".

Finally, we discuss the crossover --at charge neutrality-- between
the transport gap due to Coulomb islands, and the gap at large $B$
values. The transport gap associated to Coulomb islands is not a
"true" gap in the density of states, i.e. states are still present
at low energy. On the contrary, the strongly insulating state
appearing for $B>1$ T seems to have a true gap, since no current can
be detected at low bias over a large $V_g$ range (Fig.
\ref{fig4}(e)). Contrary to the transport gap associated to Coulomb
blockade, this gap increases with increasing $B$ (see Fig.
\ref{fig3}(a) and Fig. \ref{fig5}(b)) and at sufficiently large $B$,
it behaves virtually identically to what is found in clean suspended
layers of larger width (see Fig.
\ref{fig2}(c))~\cite{Du09,Bolotin09}. Note that, as observed in Ref.
13, the correlation induced gap  in Fig. \ref{fig4}(e) is asymmetric
around the charge neutrality point, i.e., electron-hole symmetry is
broken in the underlying electronic state .

We conclude by comparing our results to those of Ref. 11, where
Tombros {\it et al.} realized a constriction by annealing a
suspended graphene flake, and observed signatures of conductance
quantization (i.e. ballistic transport) at $B=0$ T, whereas at $B=0$
T we observe carrier localization. What appears to be mainly
responsible for the different behavior is the device length. In Ref.
11 the length of the constriction is comparable to (or shorter than)
the width; in all devices imaged by us, the length of the structures
is much longer than their width. Edge scattering is therefore much
more severe, resulting in electron localization and preventing
ballistic transport. It is clear that systematic investigations of
suspended devices as a function of their width and length are worth
pursuing, to explore the properties of confined Dirac electrons in
different transport regimes. Such a study would provide information
on the mechanisms of edge scattering and on the nature of edge
disorder, two virtually unexplored aspects of graphene
nano-structures. A difficulty is the lack of control of the
dimensions of the structures that are created by current annealing,
and it will be crucial to make the fabrication process more
controllable (e.g., by "seeding" structural defects in the graphene
layer before suspension, where annealing-induced cleavage occurs
preferentially). Even without this control, our results show that
nano-ribbons generated by annealing suspended graphene enable the
observation of new and interesting phenomena.

We acknowledges N. Tombros for discussions about device fabrication
and A. Ferreira for technical support. Financial support from the
NCCRs MANEP, and QSIT, and from SNF is  gratefully acknowledged.


\begin{thebibliography}{}
\bibitem{note1} It has not been possible to image all the
nano-ribbons produced by annealing, because most devices broke in
the course of the experiments, which gives an uncertainty about
their dimensions. In the cases that we could image, the nano-ribbon
length was comparable to the contact separation (see Fig.
\ref{fig2}(a)). For the nano-ribbon whose data we show here the
width $W$ can be estimated from the results of the measurements.
Quantum Hall plateaus starting at 2-2.5 T indicates that $W>100$ nm,
since otherwise the edge states at opposite edges would overlap,
preventing the observation of conductance quantization. By comparing
the charging energy ($\approx$ 10 meV, see Fig. \ref{fig2}(d)) to
data measured in ribbons on SiO$_2$ --and given that the dielectric
constant of SiO$_2$ partially screens Coulomb interactions-- we also
estimate that $W<200$ nm.
\end{thebibliography}
\end{document}